\documentclass[epj,twocolumn]{webofc}
\usepackage[varg]{txfonts}   
\woctitle{Flavor changing and conserving processes}
\newcommand{\lepton}{l}
\newcommand{\electron}{e}
\newcommand{\muon}{\mu}

\newcommand{\aleptonhlo}{a_{\lepton}^{\mathrm{hlo}}}
\newcommand{\aleptonbarhlo}{a_{\bar{\lepton}}^{\mathrm{hlo}}}
\newcommand{\fermi}{\,\mathrm{fm}}
\newcommand{\gev}{\,\mathrm{GeV}}
\newcommand{\mev}{\,\mathrm{MeV}}
\newcommand{\mpi}{m_{\pi}}
\newcommand{\mps}{m_{PS}}
\newcommand{\mlepton}{m_{\lepton}}
\newcommand{\mcr}{\ensuremath{m_{\mathrm{cr}}}}

\newcommand{\alphaQED}{\alpha_{\mathrm{QED}}}
\newcommand{\munu}{{ \mu\nu}}
\newcommand{\brackets}[1]{\langle #1 \rangle}
\newcommand{\epow}[1]{\mathrm{e}^{#1}}

\newcommand{\refeq}[1]{(\ref{#1})}
\newcommand{\reffig}[1]{[\ref{#1}]}
\newcommand{\reftab}[1]{\{\ref{#1}\}}

\newcommand{\Saction}{\mathcal{S}}
\newcommand{\nablaf}{\nabla^f}
\newcommand{\nablab}{\nabla^b}
\newcommand{\chibar}{\bar{\chi}}
\newcommand{\psibar}{\bar{\psi}}
\newcommand{\gammafive}{\gamma_{5}}
\newcommand{\order}[1]{\mathcal{O}\left(#1\right)}
\newcommand{\hmu}{\hat{\mu}}
\newcommand{\hnu}{\hat{\nu}}
\newcommand{\Qhat}{\hat{Q}}
\newcommand{\Qtilde}{\tilde{Q}}
\newcommand{\Real}[1]{\mathrm{Re}\left( #1 \right)}

\begin{document}
\title{ Leading-order hadronic contributions to the lepton anomalous magnetic moments from the lattice }
%
%

\author{
  Florian Burger\inst{1}
\and
  Xu Feng \inst{2}
\and
  Karl Jansen \inst{4}
\and
  Marcus Petschlies\inst{5}\fnsep\thanks{\email{marcus.petschlies@hiskp.uni-bonn.de}}
\and
  Grit Pientka \inst{3}
\and
  Dru B. Renner\inst{6}
}

\institute{
  OakLabs GmbH, Neuendorfstr. 20B, D-16761 Hennigsdorf, Germany
\and
  Columbia University, Department of Physics, 538 West 120th Street,  704 Pupin Hall, New York, NY 10027, USA
\and
  Humboldt-Universit\"at zu Berlin, Institut f\"ur Physik, Newtonstr. 15, D-12489 Berlin, Germany 
\and
  NIC, DESY, Platanenallee 6, D-15738 Zeuthen, Germany
\and
  Rheinische Friedrich-Wilhelms-Universit\"at Bonn,
  Institut f\"ur Strahlen- und Kernphysik,
  Nu{\ss}allee 14-16, D-53115 Bonn, Germany
\and
  Los Alamos National Laboratory (LANL),
  P.O. Box 1663, 
  Los Alamos, NM 87545,
  USA 
}

\noindent \quad \\ DESY 15-206

\abstract{%
  The hadronic leading-order (hlo) contribution to the lepton anomalous magnetic moments $\aleptonhlo$ of the Standard Model leptons still accounts for the dominant source of the 
  uncertainty of the Standard Model estimates. We present the results of an investigation of the hadronic leading order anomalous magnetic moments of the electron, muon and tau
  lepton from first principles in twisted mass lattice QCD. With lattice data for multiple pion masses in the range $230 \mev\lesssim \mps \lesssim 490\mev$, multiple lattice
  volumes and three lattice spacings we perform the extrapolation to the continuum and to the physical pion mass and check for all systematic uncertainties in the lattice calculation.
  As a result we calculate $\aleptonhlo$ for the three Standard Model leptons with controlled statistical and systematic error in agreement with phenomenological determinations
  using dispersion relations and experimental data. 
  In addition, we also give a first estimate of the hadronic leading order anomalous magnetic moments from simulations directly at the physical value of the pion mass.
}
\maketitle
\section{Introduction}
\label{sec:1}
Lattice QCD offers the opportunity to study hadronic contributions to electroweak observables from first principles. This conference contribution focuses on the hadronic leading order (in the electroweak couplings)
contribution to the anomalous magnetic moments $\aleptonhlo$ of the Standard Model leptons electron, muon and tau. Beyond these important quantities the calculation shown here can be adapted to a broad range of applications for a variety of quantities
such as the next-to-leading order vacuum-polarization type contributions to the lepton anomalous magnetic moments, the hadronic running of the electroweak couplings  $\alphaQED$ and $\sin^2\left( \theta \right)$, the Adler
function or $\Lambda_\mathrm{QCD}$ amongst others \cite{Renner:2012fa}.

As a crucial prerequisite, the nonperturbative hadron dynamics entering all these quantities can be calculated entirely in Euclidean spacetime. This makes them accessible to ab-initio lattice calculations, which simulate the strong interaction
of quarks and gluons on a Euclidean spacetime lattice. For the muon anomalous magnetic moment in particular, this has been worked out in the pioneering work by Blum \cite{Blum:2002ii}. On the other hand, 
the regularization of QCD by the finite lattice volume $L^3$, the lattice spacing $a$ and the simulation at unphysically large pion masses introduce a systematic dependence on the infrared and ultraviolet cutoff, $1/L$ and $1/a$, respectively,
and the light pseudoscalar mass, which need to be extrapolated with care to the corresponding limits, $L \to \infty$, $a \to 0$ and $\mps \to \mpi$.

The simulation effort by the European Twisted Mass Collaboration (ETMC) has matured to a state, at which we can calculate the above mentioned quantities for multiple lattice volumes, lattice spacings and pion masses close to or even
at the physical point with contributions from up, down, strange and charm quarks. Within this framework we are now able to reach for high-precision results from lattice QCD with control over systematic uncertainties.
In this contribution we report our progress for the hadronic leading order contribution to the anomalous magnetic moments of the Standard Model leptons towards the goal of reaching the current precision obtained in phenomenological
analyses using experimental data.

\begin{figure}[htpb]
  \centering
  \includegraphics[width=5cm,clip]{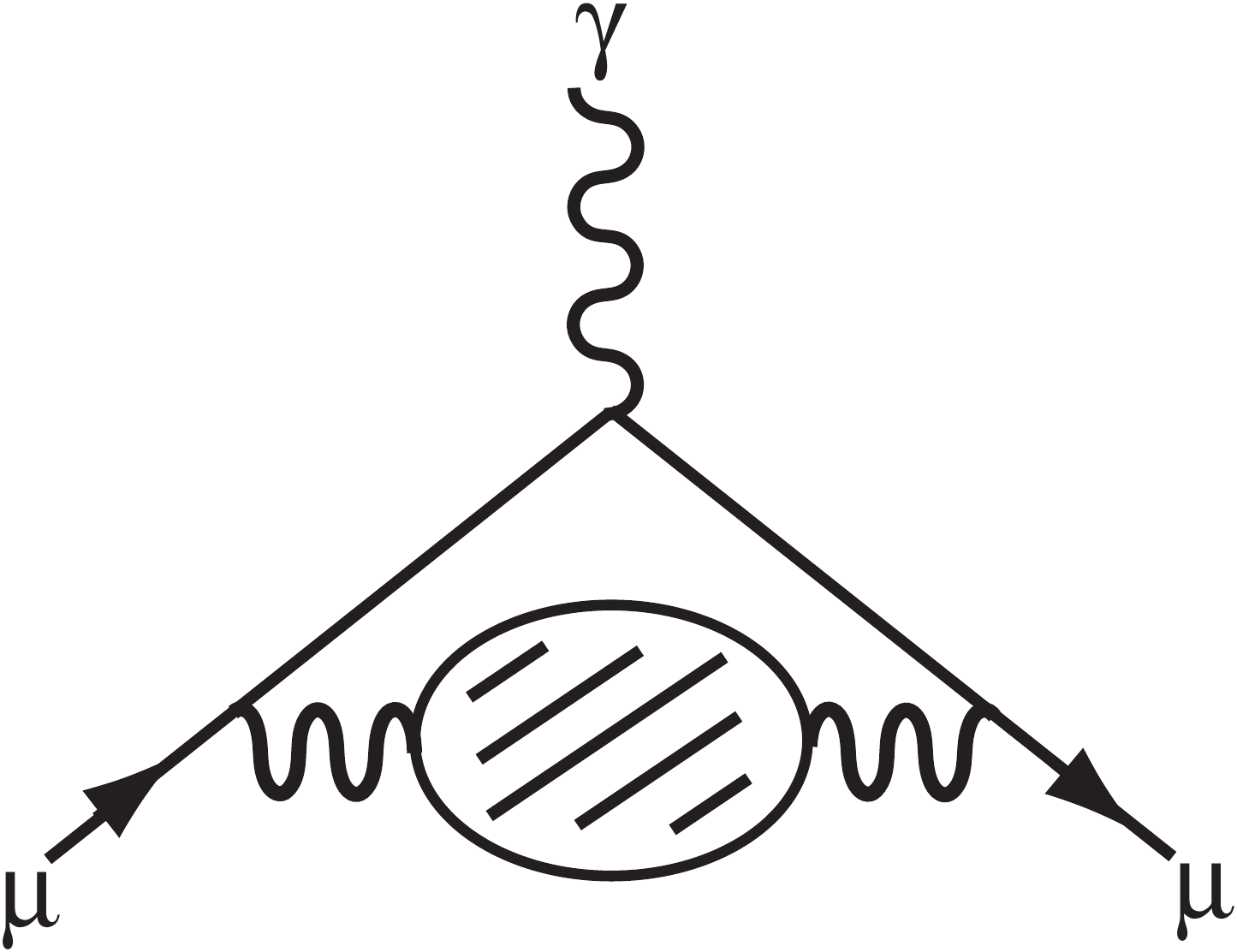}
  \caption{Hadronic leading order correction to the muon-photon-vertex}
  \label{fig:1}
\end{figure}

The essential ingredient for this calculation is the hadronic vacuum polarization tensor and function
\begin{align}
  \Pi^{AB}_\munu(Q) &= \int\limits\,d^4x\,\brackets{J^A_\mu(x)\,J^B_\nu(y)}\,\epow{iQ(x-y)} \nonumber\\
  &= \left( Q_\mu\,Q_\nu - \delta_\munu\,Q^2 \right)\,\Pi^{AB}(Q^2)\,.
  \label{eq:vp_tensor}
\end{align}
determined from lattice QCD.
Here $J^A_\mu$ denotes a flavor diagonal isospin component of the vector current or a linear combination of such components. In particular $\aleptonhlo$ requires the vacuum polarization function from the 2-point correlator
of electromagnetic currents, $A = \gamma$, as part of the integrand for the QED-1-loop integral (cf. figure \reffig{fig:1}) \cite{Blum:2002ii}
\begin{align}
  \aleptonhlo &= 4\,\alpha^2\,\int\limits\,\frac{dQ^2}{Q^2}\, \Pi^{\gamma\gamma}_R(Q^2)\,w\left(Q^2/\mlepton^2\right)\,.
  \label{eq:aleptonhlo_integral}
\end{align}
with the subtracted vacuum polarization function
\begin{align*}
  \Pi^{\gamma\gamma}_R(Q^2) &= \Pi^{\gamma\gamma}(Q^2) - \Pi^{\gamma\gamma} (0)\,.
\end{align*}

The muon anomalous magnetic moment has been determined both experimentally and theoretically to an uncertainty of approximately 0.5 ppm with a discrepancy of 2 to 4 standard deviations between the experimental and
the Standard Model value \cite{Brambilla:2014jmp}
\begin{align*}
  a_\mu(\mathrm{EXP}) &= 11 659 208.9\,(6.3) \cdot 10^{-10}\,,\\
  a_\mu(\mathrm{SM})  &= 11 659 180.4\,(4.2)\,(2.6) \cdot 10^{-10}\,.
\end{align*}
$a_\mu(\mathrm{SM})$ is not a pure Standard Model number, though, since its extraction partly relies on experimental data and some amount of modeling.
It has thus become a focal point of numerous studies in lattice QCD with an enhanced effort to investigate the systematic uncertainties which arise naturally in the framework of a lattice QCD calculation, such as
the finite lattice volume.

Though not immediately apparent from the definition in equation \refeq{eq:aleptonhlo_integral}, the extension of the lattice QCD calculation to the electron and tau hadronic leading order anomalous magnetic moments is 
a non-trivial step. From the practical point of view the calculation is universal for all three leptons, since the specific lepton property only enters via the lepton mass in the analytically known weight function
$w\left( Q^2/\mlepton^2 \right)$ in the integrand. But the ratios of the lepton masses $m_\electron^2 : m_\muon^2 : m_\tau ^2 \approx  1 : 4\cdot 10^4 : 10^7$ extending over 7 orders of magnitude imply a significantly
different weighting of the momentum dependence of the polarization function in the momentum integration for the individual leptons. As a consequence the individual lattice artifacts will have significantly different bearings on $\aleptonhlo$,
in each case, which we discuss in more detail below.

\section{Twisted mass lattice QCD calculation}
\label{sec:2}
We use the twisted-mass lattice regularization of QCD \cite{Frezzotti:2003ni} for a mass degenerate light quark doublet $\chi_l = (u,d)^T$ and a non-degenerate heavy quark doublet $\chi_h = (c,s)^T$
together with the Iwasaki gauge action \cite{Iwasaki:1985we}.
The sea-quark action is given by
\begin{align}
  \Saction_\mathrm{sea} &= \sum\limits_{x} \,\chibar_l(x)\,\left( D_W + i\mu_l\,\gammafive\,\tau^3 \right)\,\chi_l(x) \nonumber\\
 &\quad  + \sum\limits_{x} \,\chibar_h(x)\,\left( D_W + i\mu_\sigma\,\gammafive\,\tau^1 + \mu_\delta\,\tau^3 \right)\,\chi_h(x)\,.
  \label{eq:tm_sea_quark_action}
\end{align}
$D_W$ is the Wilson-Dirac operator
\begin{align}
  D_W &= \frac{1}{2}\,\gamma_\mu\,\left( \nablaf_\mu + \nablab_\mu\right) - \left( \frac{ar}{2}\,\nablab_\mu\,\nablaf_\mu - \mcr \right) + \left( m_0 + \mcr \right)
  \label{eq:Wilson_Dirac_operator}
\end{align} 
where $\nablaf,\,\nablab$ denote the gauge-covariant, discrete forward and backward derivative, $m_0$ the bare quark mass and $\tau^{1,2,3}$ the Pauli matrices.
The twisted mass term for the heavy quark doublet $i\gammafive\,\tau^1\,\mu_\sigma$ 
breaks the vector part of isospin symmetry such that the flavor diagonal isovector currents, and in particular also the electromagnetic current, are no longer conserved. We thus
use a mixed action ansatz with a valence quark action, that has improved isospin symmetry \cite{Osterwalder:1977pc,Frezzotti:2004wz,Burger:2013jya},
\begin{align}
  \Saction_\mathrm{val} &= \sum\limits_{f = l,s,c}\,\sum\limits_{x}\,\psibar_f(x)\,\left( D_W + i\mu_f\,\gammafive\,\tau^3 \right)\,\psi_f(x)
  \label{eq:tm_valence_quark_action} \\
  \psi_f &= \left( \psi^+_f,\,\psi^-_f \right)^T\,.
\end{align}
Using this mixed action on the lattice has been shown to guarantee the automatic absence of lattice artifacts of $\order{a}$ in the continuum limit of physical, on-shell observables \cite{Frezzotti:2004wz}
by solely tuning the bare quark mass $m_0$ to its critical value $m_0 \to \mcr$ without any further improvement of lattice action or operators.
This property remains true for the hadronic vacuum polarization function $\Pi^{\gamma\gamma}$ 
in the presence of off-shell, short distance contributions in equation \refeq{eq:vp_tensor} \cite{Burger:2014ada}. The renormalized strange and charm valence quark masses in the mixed action
are fixed by requiring $2 m_K^2 - \mps^2$ and the $D$-meson mass calculated in the valence sector to attain their physical value.

\begin{table}
\centering
\caption{Gauge field ensembles generated by the European Twisted Mass Collaboration with $N_f=2+1+1$ dynamical quarks \cite{Baron:2010bv,Baron:2010th}
  and with $N_f=2$ dynamical quark flavors at physical pion mass \cite{Abdel-Rehim:2013yaa,Abdel-Rehim:2014nka} (last row).}
\label{tab:etmc_ensembles}
  \begin{tabular}{|c | c c c c|}
  \hline
  & & & &\vspace{-0.40cm} \\
  Ensemble & $a[{\rm fm}]$ & $m_{PS}$[MeV] & $L$[fm] & $m_{PS}\cdot L$\\
  & & & &\vspace{-0.40cm} \\
  \hline \hline
  & & & &\vspace{-0.40cm} \\
  D15.48   & $0.061$ & 227 & 2.9  & 3.3 \\
  D30.48   & $0.061$ & 318 & 2.9  & 4.7 \\
  D45.32sc & $0.061$ & 387 & 1.9  & 3.7 \\
  \hline
  & & & & \vspace{-0.40cm} \\
  B25.32t  & $0.078$ & 274 &  2.5 & 3.5 \\
  B35.32   & $0.078$ & 319 &  2.5 & 4.0 \\
  B35.48   & $0.078$ & 314 &  3.7 & 5.9 \\
  B55.32   & $0.078$ & 393 &  2.5 & 5.0 \\
  B75.32   & $0.078$ & 456 &  2.5 & 5.8 \\
  B85.24   & $0.078$ & 491 &  1.9 & 4.7 \\
  & & & &\vspace{-0.40cm} \\
  \hline
  & & & &\vspace{-0.40cm} \\
  A30.32   & $0.086$ & 283 & 2.8 & 4.0 \\
  A40.32   & $0.086$ & 323 & 2.8 & 4.6 \\
  A50.32   & $0.086$ & 361 & 2.8 & 5.1 \\
  & & & &\vspace{-0.40cm} \\
  \hline
  & & & &\vspace{-0.40cm} \\
  cA2.09.48& $0.091$ & $\approx \mpi$ & 4.4 & 3.0 \\
  \hline
  \end{tabular}
\end{table}
The gauge field ensembles with $N_f=2+1+1$ dynamical quarks produced by the ETMC and used for this calculation are listed in table \reftab{tab:etmc_ensembles}. Details about their production and properties
are given in references \cite{Baron:2010bv,Baron:2010th}. Additionally, we consider one ETMC ensemble at the physical pion mass \cite{Abdel-Rehim:2013yaa,Abdel-Rehim:2014nka,Abdel-Rehim:2015pwa},
whose parameters are listed in the last row of the table.

For the valence quark action eq. \refeq{eq:tm_valence_quark_action} the electromagnetic current on the lattice follows as the Noether current from the vector flavor variation
\begin{align}
  \delta\,\psi(x) &= i Q^\mathrm{em}\,\psi (x)\,,\quad 
  \delta\,\psibar(x) = -i \psibar(x)\,Q^\mathrm{em}
  \label{eq:qem_vector_flavor_variation}
\end{align}
generated by the electromagnetic charge matrix $Q^\mathrm{em} = \mathrm{diag}\left( +2/3,\,-1/3,\ldots \right)$ acting in flavor space.
Eq. \refeq{eq:qem_vector_flavor_variation} defines the electromagnetic current as the usual sum of single-flavor quark currents
\begin{align*}
  J^\gamma_\mu &= \frac{2}{3}\, J^u_\mu -\frac{1}{3}\, J^d_\mu + \frac{2}{3}\, J^c_\mu -\frac{1}{3}\, J^s_\mu
\end{align*}
and the lattice vacuum polarization tensor
\begin{align}
  \Pi^{\gamma\gamma}_\munu(x,y) &= \brackets{J^\gamma_\mu(x)\,J^\gamma_\nu(y)} - a^{-3}\,\delta_{x,y}\,\delta_\munu\,\brackets{S_\nu(y)}
  \label{eq:lattice_vacuum_polarization_tensor}
\end{align}
as the 2-point current correlator with a lattice contact term. The tensor fulfills the lattice version of exact transversality at non-zero lattice spacing
\begin{align}
  \partial^b_\mu\,\Pi^{\gamma\gamma}_\munu(x,y) &= 0
  \label{eq:vp_transverse_position_space}
\end{align}
with the discrete backward partial derivative $\partial^b_\mu$.

The discrete 4-dimensional Fourier transform translates this transversality to momentum space
\begin{align*}
  \Pi_\munu(Q) &= \sum\limits_{x}\,\Pi_\munu(x,y)\,\epow{iQ(x+a\hmu/2 - y - a\hnu/2)} \\
  \Qhat_\mu\,\Pi_\munu(Q) &= 0\,,\quad \Qhat_\mu = \frac{2}{a}\,\sin\left(a Q_\mu/2 \right)
\end{align*}
with $Q_i \in 2\pi/L \cdot \left\{ 0,\ldots,L/a-1 \right\}$ and $Q_0 \in 2\pi/T \cdot \left\{ 0,\ldots,T/a-1 \right\}$ for the spatial and temporal momentum components.
We extract the polarization function according to
\begin{align}
  \Pi^{\gamma\gamma}\left( \Qhat^2 \right) &=
  \frac{1}{\#\,\left[ \Qhat^2 \right]}\,
    \sum\limits_{Q\in \left[ \Qhat^2 \right]}\,
    \Real{ \sum\limits_{\mu,\nu}\, P_\munu(Q)\,\Pi^{\gamma\gamma}_\munu(Q) } \nonumber\\
&\qquad    \times \left(  \sum\limits_{\mu,\nu}\,P_\munu(Q)\,P_\munu(Q)  \right)^{-1}
  \,,
  \label{eq:lattice_vacuum_polarization_function} \\
  P_\munu(Q) &= \Qhat_\mu\,\Qhat_\nu - \delta_\munu\,\Qhat^2\,.\nonumber
\end{align}
The definition in eq. \refeq{eq:lattice_vacuum_polarization_function} includes the projection with $P_\munu$, which is non-trivial for the index $\nu$, and
discarding the imaginary part, which is a lattice artifact due to flavor symmetry breaking. With a $\left[ \Qhat^2 \right]$ we denote the set of all lattice
momenta $Q$ with the same $\Qhat^2$ and with $\#\,\left[ \Qhat^2 \right]$ the number of such momenta.

\begin{figure}
\centering
\includegraphics[width=7cm,clip]{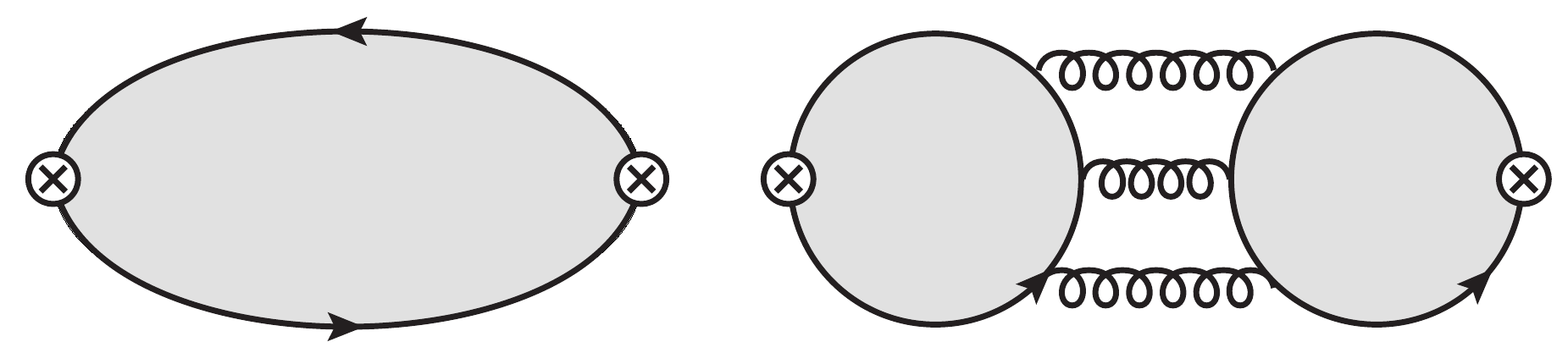}
\caption{Diagrammatic representation of the quark-connected (left) and (exemplary) quark-disconnected contribution to the vacuum polarization;
black lines denote valence quark loops and gray shading represents full, non-perturbative QCD interaction.}
\label{fig:2}
\end{figure}

The vacuum polarization tensor and function decompose into a dominant quark-connected and a small quark-disconnected contribution, which are illustrated by the left-hand
and right-hand diagram in figure \reffig{fig:2}, respectively.

\begin{figure*}
\centering
\includegraphics[width=0.48\textwidth]{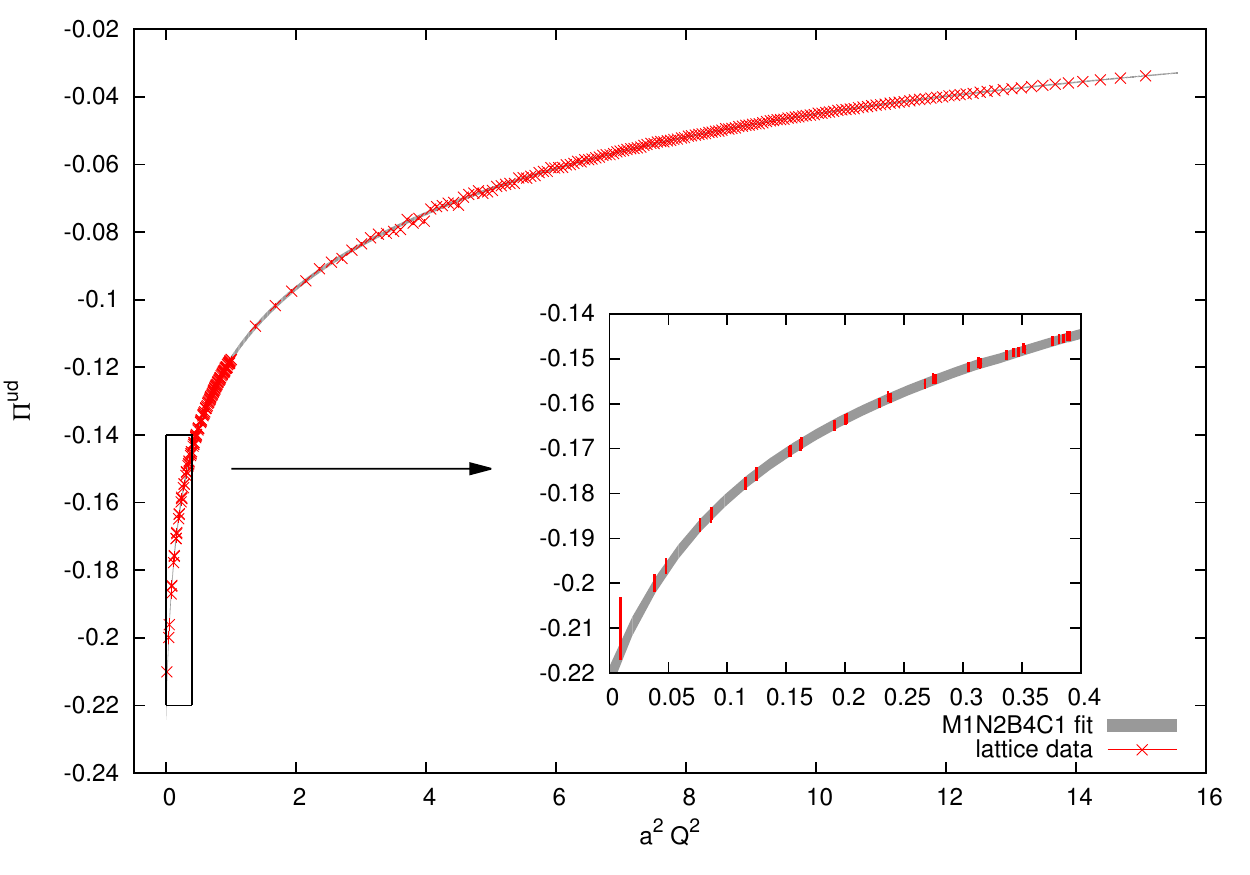}
\includegraphics[width=0.48\textwidth]{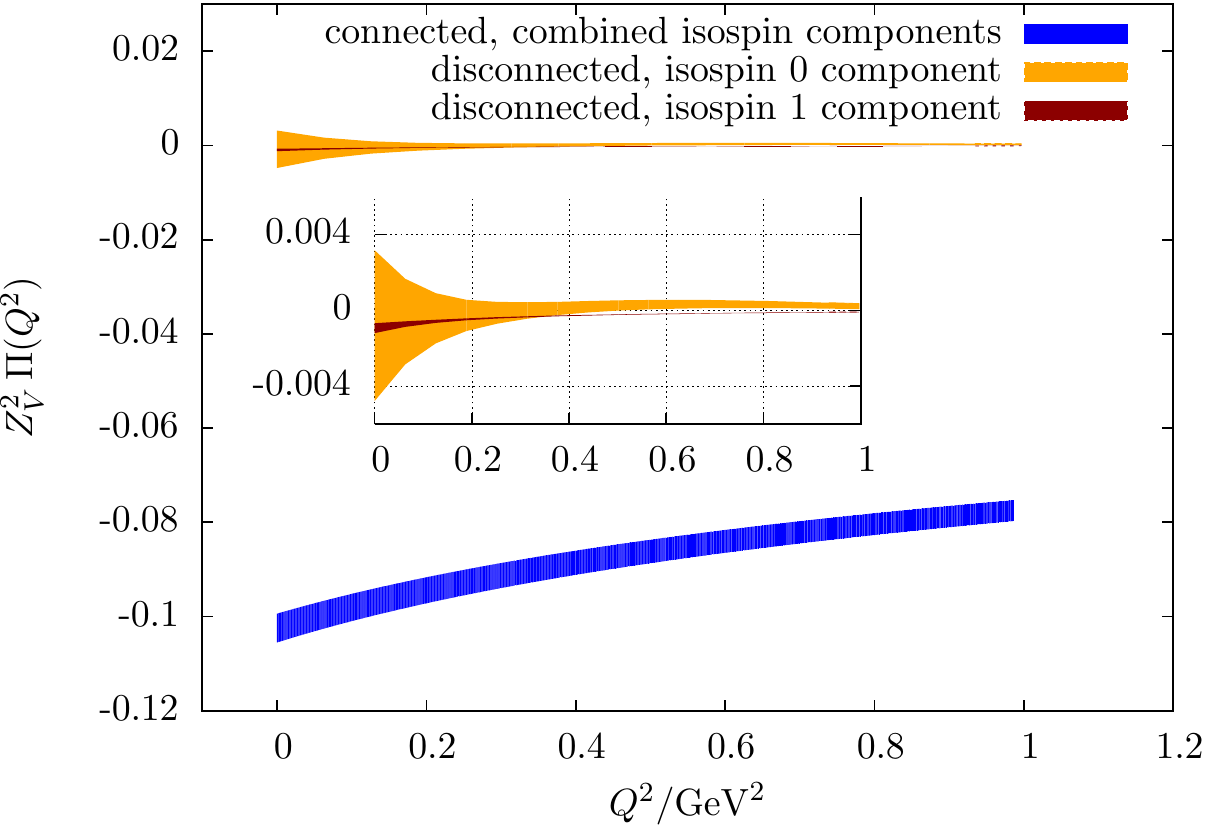}
\caption{Left: quark-connected contribution from up and down quark to the hadronic vacuum polarization function; the data shown belongs to ensemble B25.32t ($\mps = 274\mev$, $L = 2.5\fermi$ and $a = 0.078\fermi$).
Right: quark-disconnected contribution from up and down quark for ensemble B55.32 ($\mps = 393\mev$, $L = 2.5\fermi$ and $a = 0.078\fermi$).The detail plots show a magnification of the small-momentum region.}
\label{fig:3}
\end{figure*}

For illustration we show exemplary data for the connected and disconnected contribution to the vacuum polarization function in the left and right panel of figure \reffig{fig:3}, respectively.
The left-hand panel shows the typical smooth momentum dependence of the connected contribution from up and down quarks for ensemble B25.32t (cf. table \reftab{tab:etmc_ensembles}) between the infrared and ultraviolet lattice cutoff 
$(2\sin(a\pi/T)/a)^2 \le \Qhat^2 \le 16 / a^2$.
The gray band in the plot shows our standard fit to the data, which will be discussed below. In the detail plot we magnify the crucial region at small momenta, where lattice data points become sparse.
The right-hand panel shows an order-of-magnitude comparison of the signal for the light-quark connected polarization function (blue band) and the disconnected contribution from the 2-point correlation of the 
isovector components of the electromagnetic current (dark-red band labeled isospin 1) and of the isoscalar components (yellow band labeled isospin 0). The data shown was obtained with particularly high
statistics for ensemble B55.32, cf. \cite{Michael:2013gka}. While the contribution from the isovector currents is a lattice artifact in the twisted mass regularization, the physically meaningful contribution from the isoscalar vector current correlator
appears about two orders of magnitude smaller compared to the connected contribution. In fact, with the presently available data we cannot resolve a statistically significant contribution from the disconnected vacuum polarization
function to $\aleptonhlo$. In the following we will thus be concerned only with the connected vacuum polarization function.

\begin{figure*}
\centering
\includegraphics[width=0.435\textwidth]{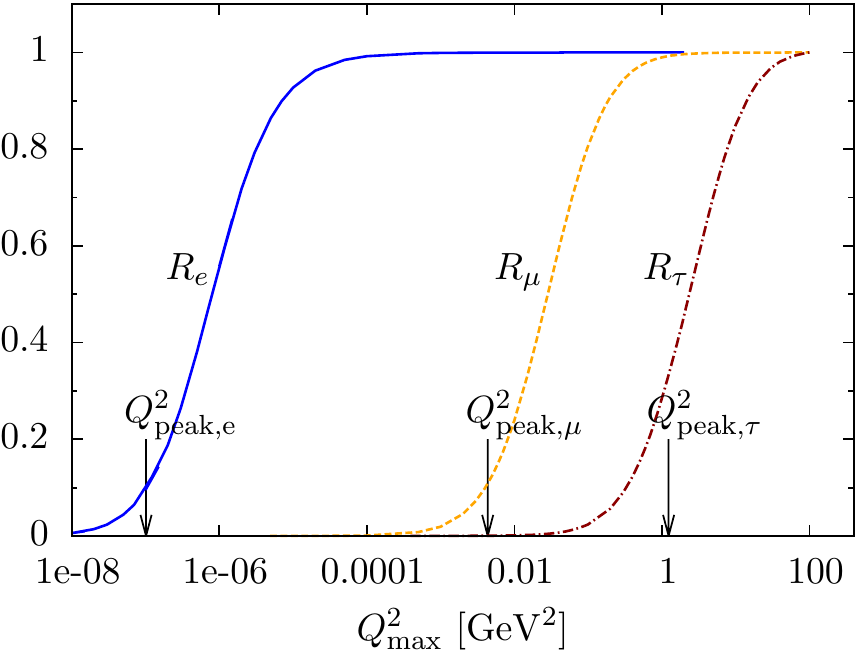}
\includegraphics[width=0.545\textwidth]{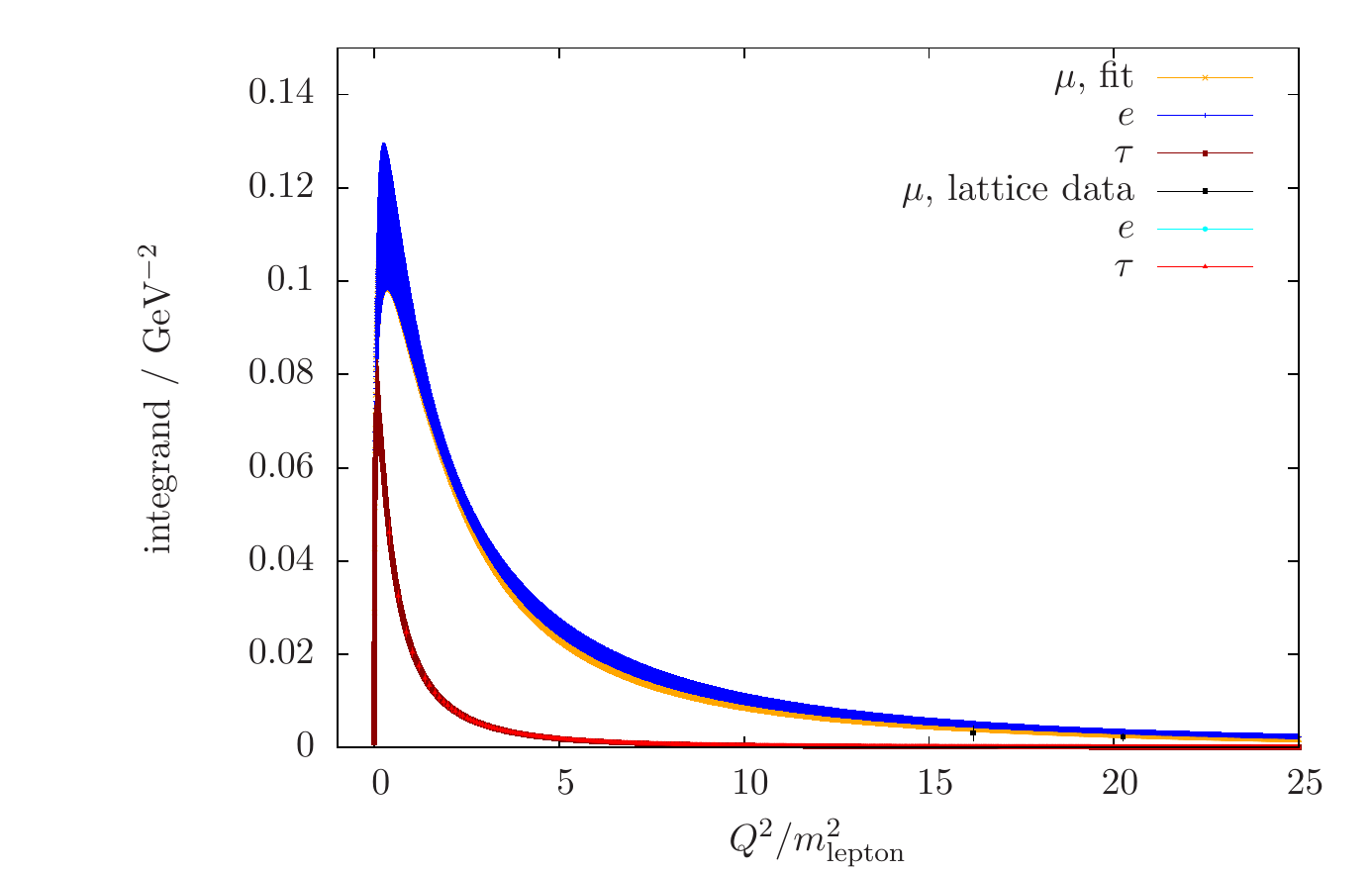}
\caption{Left: saturation of the integral defining $\aleptonhlo$ for $\lepton = \electron,\,\muon,\,\tau$; the data shown
corresponds to ensemble B55.32. Right: the full integrand for ensemble D15.48 in a universal plot for all three leptons.}
\label{fig:4}       
\end{figure*}

\paragraph{Momentum dependence of the polarization function}
The lattice calculation generically provides us with data points for the vacuum polarization function for a finite set of discrete, nonzero momenta. According to equation \refeq{eq:aleptonhlo_integral}
the subtracted polarization function enters the integrand and the integral extends beneath the lowest non-zero lattice momentum to zero. We thus need to extrapolate the existing lattice data to all
$Q^2 < Q^2_\mathrm{min} = 4/a^2\sin^2(\pi/L)$. Secondly, the infinite upper integration limit is relaxed to the lattice cut-off $Q^2_\mathrm{max} = 16/a^2$. To appreciate the significance of these two issues
we display the integral saturation in the left-hand panel of figure \reffig{fig:4}: shown is the ratio
\begin{align*}
  R_\lepton\left( Q^2_\mathrm{max} \right) &= \frac{\aleptonhlo\left( Q^2_\mathrm{max} \right)}{\aleptonhlo\left( 100\gev^2 \right)}\,.
\end{align*}
of $\aleptonhlo$ integrated from zero up to $Q^1_\mathrm{max}$ over the $\aleptonhlo$ integrated from zero up to $100\gev^2$.
The plot shows, that the upper integration limit $Q^2_\mathrm{max}$ on the lattice is not problematic, since for electron and muon the integral saturates long before reaching the lattice momentum
cut-off and also for the tau the change of the integral value with $Q^2_\mathrm{max} \lesssim 100\gev^2$ is insignificant compared to the statistical and other sources of systematic uncertainties.

In the same plot $Q^2_{\mathrm{peak},\lepton}$ marks the maximum of the weight function $w\left( Q^2/m_\lepton^2 \right)$ for each lepton, which roughly locates the main support area for the integral.
As is well known, for the electron and the muon this main support region is far below the lowest reachable lattice momentum for currently available lattice volumes. In particular $Q^2_\mathrm{min} \gtrsim 0.1\gev^2$ for
our ensembles in table \reftab{tab:etmc_ensembles}. In the extreme case of the electron, $\aleptonhlo$ is essentially determined by the $d\Pi^{\gamma\gamma}/dQ^2(0)$ alone. The right panel in figure \reffig{fig:4} shows
the location of lattice data in comparison to the fitted integrand exemplarily for ensemble D15.48. While for the electron all lattice data is outside the displayed momentum range, for the muon the (black) lattice data points start at $Q^2/m_\muon^2 \approx 16$.
For the tau (red) lattice data stops short of the peak on the right-hand slope of the integrand (brown band).

To fit the momentum dependence of the polarization function numerous fit ans\"atze can be constructed based on the analyticity and spectral properties of the polarization function, resonances and other features.
In particular fit functions of the Pad\'e type have been shown to be a suitable set \cite{Aubin:2012me,Golterman:2013vca,Golterman:2014ksa} of functions with correct convergence properties to reproduce the
analytic structure of $\Pi^{\gamma\gamma}$ in the limit of infinite volume and data precision. We here use the $MNBC$ fit ansatz, which is conceptually close to the Pad\'e type fits. 
Since at this point we consider only the connected
part of $\Pi^{\gamma\gamma}$ we can decompose it immediately into single-quark-flavor components
\begin{align*}
  \Pi^{\gamma\gamma} &= \frac{5}{9}\, \Pi^{u+d} + \frac{1}{9}\, \Pi^s + \frac{4}{9}\, \Pi^c\,.
\end{align*}
Further on, we decompose the integration interval in two regions: low momentum $0 \le Q^2 \le Q^2_\mathrm{match}$ and high momentum $Q^2_\mathrm{match} < Q^2 \le Q^2_\mathrm{max}$.
Then for each individual quark flavor $f = u+d,\,s,\,c$ we use the family of fit functions
\begin{align}
  \Pi^f_\mathrm{low}\left( Q^2 \right) &= \sum\limits_{j=1}^M\,g_j^{f 2}\,\frac{m_j^{f 2}}{m_j^{f 2} + Q^2} + \sum\limits_{j=0}^{N-1}\,a^f_j\,\left( Q^2 \right)^j
  \label{eq:pi_fit_fucntion_low}\,,\\
  \Pi^f_\mathrm{high}\left( Q^2 \right) &= \log\left( Q^2\, \right)\sum\limits_{j=0}^{B-1}\,b^f_j\,\left( Q^2 \right)^j + \sum\limits_{j=0}^{C-1}\,c^f_j\,\left( Q^2 \right)^j
  \label{eq:pi_fit_fucntion_high}
\end{align}
and the full fit function is given by the sum of low- and high-momentum part weighted by Heaviside step functions
\begin{align}
  \Pi^f\left( Q^2 \right) &= \left( 1 - \Theta\left( Q^2 - Q^2_\mathrm{match} \right) \right) \,\Pi^f_\mathrm{low}\left( Q^2 \right) + \nonumber\\
  & +\quad \Theta\left( Q^2 - Q^2_\mathrm{match} \right) \,\Pi^f_\mathrm{high}\left( Q^2 \right)\,.
  \label{eq:pi_fit_function_full}
\end{align}
The set of pairs $\left\{ \left( g^f_i,\,m^f_i \right)\,| \, i=1,\ldots,M\right\}$ of couplings and masses are determined from the corresponding connected, time-dependent single-quark-flavor vector current
2-point function at zero momentum. The remaining polynomial in $\Pi_\mathrm{low}$ can be understood as the contributions from poles with position above $Q^2_\mathrm{match}$; for these terms
we can safely use a Taylor expansion around the origin. We emphasize, that $\Pi_\mathrm{high}$ is only relevant for interpolation of the lattice data.

Finite volume effects related to the infrared momentum cut-off and the dependence on the specific fit function are then probed by varying the number of parameters $M,N,B,C$. In practice,
we can resolve $M =1$ or 2 poles with $N =1,\,2$ polynomial terms. Beyond that the fit parameters both from the multi-exponential fit of the vector current 2-point function as well as the
fit of the vacuum polarization function become highly correlated. Moreover, the variation of $B$ and $C$ leads to insignificant
variations of the integral value.

\paragraph{External scales problem and modified observables} 

As a peculiarity of the lattice QCD regularization, which excludes the dynamics of leptons and photons, the lepton mass in equation \refeq{eq:aleptonhlo_integral} enters as an external scale, which is
not related to the lattice parameters. In particular, there is no inherent value for the lepton mass in lattice units, such that the lattice spacing is explicitly required to convert the lepton mass
to lattice units. By this procedure, however, the originally dimensionless $\aleptonhlo$ becomes effectively dependent on the lattice QCD scale setting, which can be monitored by considering
its effective dimension at fixed bare coupling \cite{Feng:2011zk,Renner:2012fa},
\begin{align*}
  \mathrm{d}_\mathrm{eff}\left[ \aleptonhlo \right] &= -\left. 
  \left( \frac{a}{\aleptonhlo} \right)\,
  \left( \frac{ \partial \aleptonhlo }{ \partial a } \right) \right|_{g_0 = \mathrm{fixed}} = \order{-2 \sim -1}
\end{align*}
The latter ranges from approximately -2 for the electron to -1 for the tau, such that $\aleptonhlo$ behaves rather like an inverse mass, $M^{\mathrm{d}_\mathrm{eff}}$, than a dimensionless
quantity. This external scale problem has been solved by introducing a new family of modified observables, $\aleptonbarhlo$ defined as follows. We denote with $\Qtilde$ the momentum in
lattice units and can rewrite the integral in equation \refeq{eq:aleptonhlo_integral} in terms of $\Qtilde$ and define for any QCD scale $H$, which can be calculated on the lattice
\begin{align}
  \aleptonbarhlo &= 4\,\alpha^2\,\int\limits\,\frac{d \Qtilde^2}{\Qtilde^2}\, \Pi^{\gamma\gamma}_R\left(\Qtilde^2\right) \,
  w\left( \frac{\Qtilde^2}{a^2 H^2} \cdot \frac{H_\mathrm{phys}^2}{\mlepton^2}\right)
  \label{eq:modifield_observable}
\end{align}
with the consistency condition
\begin{align*}
  \lim\limits_{\mps \to \mpi} H(\mps) &= H_\mathrm{phys}(\mpi)
\end{align*}
such that at the physical point the modified observable agrees with the standard definition for each choice of a consistent $H$
\begin{align}
  \lim\limits_{\mps \to \mpi} \,\aleptonbarhlo(\mps) &= \aleptonhlo(\mpi)\,.
  \label{eq:modifield_observable_consistency}
\end{align}
The introduction of a QCD-scale $H$ redefines the lepton mass in the lattice calculation
\begin{align*}
  m_{\bar{\lepton}} &= m_\lepton / H_\mathrm{phys} \cdot H\,,
\end{align*}
such that the modified lepton mass becomes dependent on the lattice scale setting and
\begin{align*}
  \mathrm{d}_\mathrm{eff}\left[ m^2_{\bar{\lepton}} \right] &= 2 \quad \mathrm{~and~} \quad
  \mathrm{d}_\mathrm{eff}\left[ \aleptonbarhlo \right] = 0\,.
\end{align*}
In practice we use the empirically motivated choice $H = m_V$, the mass of
the ground state in the light vector channel. This exploits the strong statistical correlation of
the light vector meson mass and $\aleptonhlo$.

\begin{figure}
\centering
\includegraphics[width=0.50\textwidth,clip]{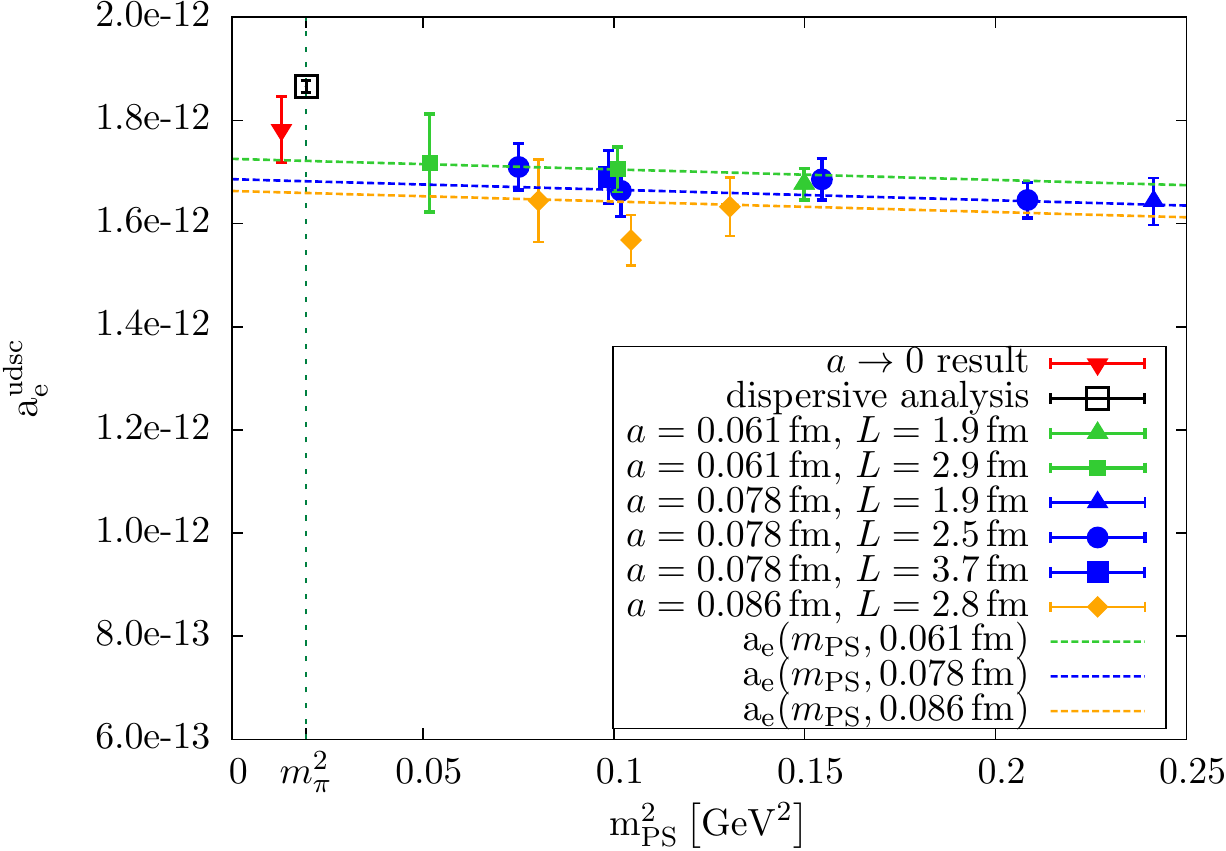}
\includegraphics[width=0.50\textwidth,clip]{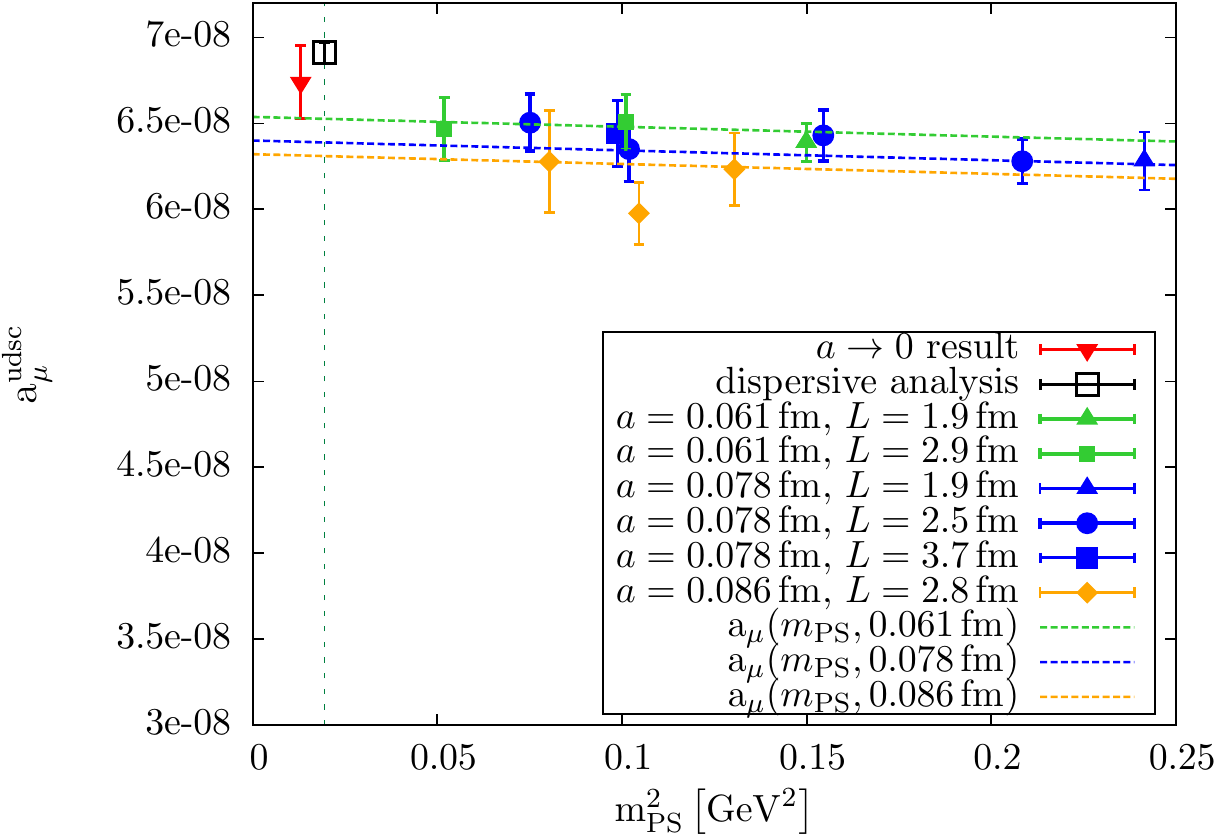}
\includegraphics[width=0.50\textwidth,clip]{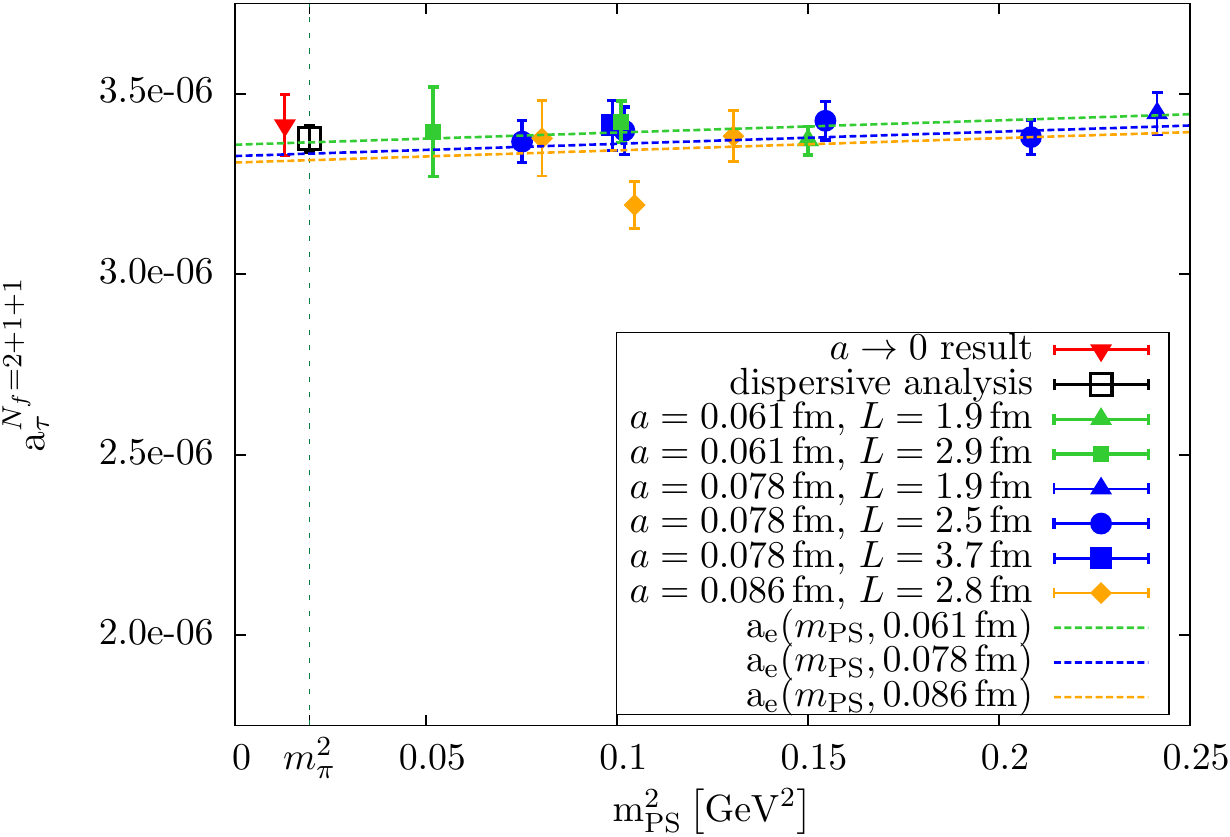}
\caption{Combined extrapolation in pion mass and lattice spacing for $\aleptonbarhlo$ for electron (top),
  muon (center) and tau (bottom) using free fit parameters $A,\,B_1,\,C$ in equation \refeq{eq:extrapolation_mps_a}.}
\label{fig:5}       
\end{figure}

\paragraph{Extrapolation and results}
To extrapolate in the light pseudoscalar mass to the physical point $\mps \to \mpi$ and to the continuum we use the ansatz
\begin{align}
  \aleptonbarhlo\left( \mps,\,a \right) &= A + B_1\,\mps^2\, \left( + B_2\,\mps^4 + \ldots \right) + C\,a^2\,.
  \label{eq:extrapolation_mps_a}
\end{align}
These extrapolations of the lattice data for $\aleptonbarhlo$ are shown in figure \reffig{fig:5} for electron (top),
muon (center) and tau lepton (bottom).
Inspection by eye of the data points strongly suggests a linear dependence of the modified observable $\aleptonbarhlo$ on the pion mass
and it turns out, that usually optimizing only $B_1$ already provides a good fit. With (at most) $B_2$ we probe for potential
non-linearity in the pion mass dependence of our lattice data, which turns out to be insignificant.
The three dashed lines in each plot show the evaluation of the fit function for the corresponding lattice spacing. The result
for zero lattice spacing and physical pion mass is given by the red downward triangle shifted slightly off to the left
from the physical pion mass marked by the vertical dashed line. 

\begin{figure}
\centering
\includegraphics[width=0.48\textwidth,clip]{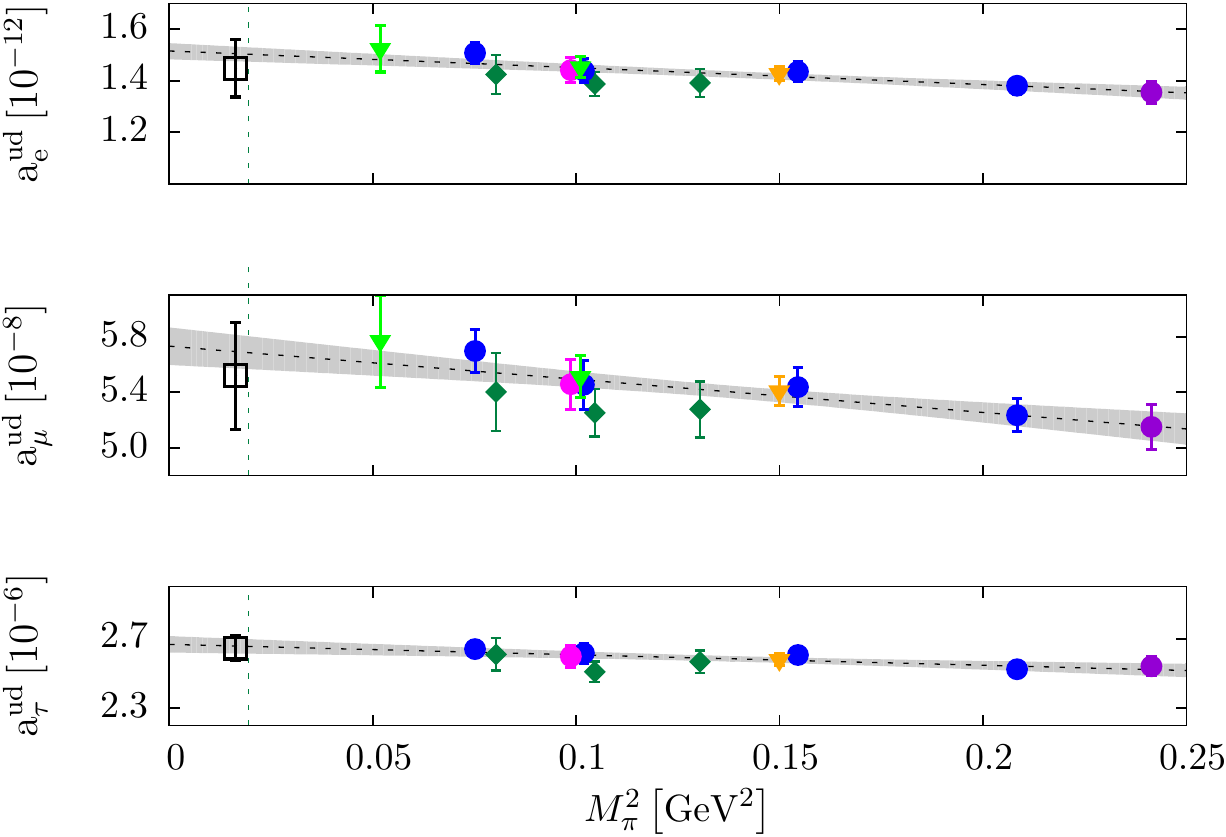}
\caption{Light-quark contribution to $\aleptonhlo$ at physical pion mass for electron (top),
  muon (center) and tau (bottom) from the standard definition with $H=1$ (cf. eq. \protect{\refeq{eq:modifield_observable}}); shown is the comparison of the $\mps^2$- extrapolated data from 
\cite{Burger:2013jya,Burger:2015oya} with the result calculated directly at the physical pion mass
\cite{Abdel-Rehim:2015pwa}.}
\label{fig:6}       
\end{figure}

Additionally, in figure \reffig{fig:6} we show the light-quark contribution (using only $\Pi^{u+d}$) for $\aleptonbarhlo$ for all three leptons: we compare
the results from the $N_f=2+1+1$ simulations, which are extrapolated to the physical pion mass (\cite{Burger:2013jya,Burger:2015oya}) with
the recent calculation on an ETMC $N_f=2$ ensemble directly at physical pion mass \cite{Abdel-Rehim:2015pwa}. The results for the simulation directly at the physical
pion mass are obtained using the standard definition of the observables (cf. eq. \refeq{eq:aleptonhlo_integral} or \refeq{eq:modifield_observable} with $H = 1$).
Within presently reachable statistical uncertainties, there is full agreement of extrapolated results and results obtained from the simulation at physical pion mass.


\paragraph{Systematic uncertainties and error budget}
We investigate comprehensively the impact of all systematic uncertainties $X$ in our lattice calculation and call the corresponding estimate $\Delta_X$. In detail we probe
\begin{enumerate}
  \item[(i)] finite size effects, $\Delta_{\mathrm{FSE}}$, by restricting the range in $\mps\cdot L$ of lattice data entering the combined continuum and chiral extrapolation to $\mps\cdot L > 3.8$
    and by comparing $\aleptonbarhlo$ for ensembles with different lattice volumes at fixed lattice spacing and pion mass;
  \item[(ii)] the impact of data from larger pion masses, $\Delta_\chi$, by restricting to lattice data in the range $\mps < 400\mev$;
  \item[(iii)] the impact $\Delta_V$ of varying the fit range for the vector meson 2-point function; 
  \item[(iv)] the dependence on the fit function $\Delta_{MNBC}$ by varying the number of parameters $M,N,B,C$ (cf. eq. \refeq{eq:pi_fit_fucntion_low}, \refeq{eq:pi_fit_fucntion_high});
  \item[(v)] the impact of disconnected contributions, $\Delta_{\mathrm{disc}}$;
  \item[(vi - viii)] the uncertainty from matching the Osterwalder-Seiler quark masses, $\Delta_\mathrm{OS}$, from varying the transition point from low- to high-momentum region, $\Delta_{Q^2_\mathrm{match}}$, and
    the uncertainty from mistuned strange and charm sea quark masses, $\Delta_{(\mu_\sigma,\mu_\delta)}$.
\end{enumerate}
Though some of these effects are likely correlated, in particular (i) and (iv), we take a conservative approach and check each uncertainty individually, thus potentially overestimating
the overall uncertainty. As a general observation, the contributions from (vi) to (viii) are negligible and we set $\Delta_{\mathrm{OS}} = \Delta_{Q^2_\mathrm{match}} = \Delta_{(\mu_\epsilon,\mu_\delta)} = 0$.
For (v) we showed above (cf. figure \reffig{fig:3}), that with presently available statistics, the impact of disconnected contributions is also negligible in our calculation, they do not lead to a statistically
significant shift of the result. Moreover, the lattice data for $\aleptonbarhlo$ in figure \reffig{fig:5} covering $3 \lesssim \mps\cdot L \lesssim 6$ and the extrapolations
indicate small finite size effects, which is confirmed again by the statistical analysis. Already above we commented on the apparently highly dominant linear dependence of the modified $\aleptonbarhlo$
on the $\mps^2$. This too leads to statistically non-discernible effects when probing for terms of higher than first order in $\mps^2$ 
with the present data. We list the estimates for the remaining two sources of systematic errors in table \reftab{tab:3}.
\begin{table}
  \caption{Estimates for systematic uncertainties $\Delta_{V}$ and $\Delta_{MNBC}$ for our lattice calculation.}
\label{tab:3}       
\centering
\begin{tabular}{|l|r r|}
  \hline
  $\aleptonhlo$ & $\Delta_V$ & $\Delta_{MNBC}$ \\
  \hline\hline
  & & \\
  $\electron$ & $0.035 \cdot 10^{-12}$ & $0.078 \cdot 10^{-12}$ \\
  $\muon$     & $0.13  \cdot 10^{- 8}$ & $0.09  \cdot 10^{- 8}$ \\
  $\tau$      & $0.046 \cdot 10^{- 6}$ & $0.032 \cdot 10^{- 6}$ \\
 \hline
\end{tabular}
\end{table}
Finally, in table \reftab{tab:2} we list the complete results for $\aleptonhlo$ for electron, muon and tau and compare to the results from phenomenological analyses based on dispersion relations
and using experimental data. Where available, we give the decomposition of the total error $(\Delta_{\mathrm{stat}})\,(\Delta_\mathrm{sys})$ into
statistical uncertainty in the first bracket and the second bracket quotes the systematic uncertainty. The systematic uncertainty
for the lattice results is obtained as $\Delta_{\mathrm{sys}} = \sqrt{\sum\limits_X\,\Delta_X^2}$.
\begin{table}
\caption{Comparison of results for $\aleptonhlo$ from $N_f=2+1+1$ lattice QCD (center) and dispersive analyses (right).}
\label{tab:2}       
\centering
{\small
\begin{tabular}{|l|r r|}
  \hline
  $\aleptonhlo$ & tmLQCD & disp. analyses \\
  \hline\hline
  & & \\
  $\electron$ & $1.782 \,(64)\,(85)\cdot 10^{-12}$ \cite{Burger:2015oya} &  $1.866 \,(10)\,(05)\cdot 10^{-12}$ \cite{Nomura:2012sb}      \\
  $\muon$     & $6.78  \,(24)\,(16)\cdot 10^{-8}$  \cite{Burger:2013jya} &  $6.91  \,(01)\,(05)\cdot 10^{-8}$  \cite{Jegerlehner:2011ti} \\
  $\tau$      & $3.41  \,( 8)\,( 6)\cdot 10^{-6}$  \cite{Burger:2015oya} &  $3.38  \,( 4)    \cdot 10^{-6}$    \cite{Eidelman:2007sb}    \\
 \hline
\end{tabular}
}
\end{table}
 
\section{Discussion and outlook}
With this contribution we report on the determination of the hadronic leading order anomalous magnetic moments of all three Standard Model leptons by the ETM Collaboration with $N_f = 2+1+1$ dynamical
quarks. The center column of table \reftab{tab:2} gives
our estimates for $\aleptonhlo$ in the continuum and at the physical pion mass with statistical and a conservatively determined overall systematic uncertainty. We find agreement with the estimates from
phenomenological analyses within presently reached uncertainties. With the lattice data available for this analysis the uncertainties from the lattice calculation are still significantly larger than
those obtained from the dispersive analyses. 
As a result of this analysis we reach a precision of roughly 6\% and 4\% for the case of the electron and muon, respectively, and
of 3\% for the tau lepton with combined statistical and systematic error.
The smaller uncertainty for the case of the tau may be expected, since the main support for the integral in equation \refeq{eq:aleptonhlo_integral} is shifted towards a region of momenta, where
lattice data for the vacuum polarization tensor is actually available. Yet we emphasize, that also in this case there is a critical dependence on the extrapolation of the vacuum polarization function
in the small-momentum region: it appears through the subtraction of $\Pi^{\gamma\gamma}(Q^2 = 0)$, and this latter value is predicted from the extrapolation of the lattice data. The determination of $a_\electron^\mathrm{hlo}$ on the lattice
is a strong test of the lattice determination of the vacuum polarization function at smallest momenta, since $a_\electron^\mathrm{hlo}$ is essentially determined by the derivative of the polarization function
at the origin, $\left. d\Pi^{\gamma\gamma}(Q^2)/dQ^2\right|_{Q^2=0}$. Thus the agreement obtained for $a_\electron^\mathrm{hlo}$ is both remarkable and encouraging. With the results from our recent determination of the
hadronic leading order contribution to the running of the electroweak couplings \cite{Burger:2015lqa}, this test has been successfully taken another step further: in this reference we demonstrate the agreement of
the renormalized vacuum polarization function determined from $N_f=2+1+1$ twisted mass lattice QCD with the dispersive relation result in a continuous interval $\left[0,\,10\gev^2  \right]$, albeit again
with somewhat larger uncertainty on the lattice result.

The use of the modified observables $\aleptonbarhlo$ (cf. equation \refeq{eq:modifield_observable}) at unphysically large pion masses and non-zero lattice spacing has proven superior to the standard definition
of $\aleptonhlo$ in equation \refeq{eq:aleptonhlo_integral} in our lattice calculation. The reason is twofold: Firstly, it is due to the introduction of the scale $H$, that the strong pion mass dependence 
of $\aleptonhlo$ is tamed to the level shown in figure \reffig{fig:5}. As a result we are able to perform a controlled chiral extrapolation with insignificant variations of the result when probing the pion mass
dependence beyond the leading, linear term. With the standard definition a controlled extrapolation from the range of pion masses we have available (cf. table \reftab{tab:etmc_ensembles}) will require some modeling
or otherwise lead to a prohibitively large uncertainty at the physical point (cf. \cite{Feng:2011zk,Renner:2012fa} for a more detailed discussion). Secondly, by an optimized choice of the scale $H$, which for us
is $H = m_V$, the light vector meson mass, the per-ensemble estimate of $\aleptonbarhlo$ benefits from a partial cancellation of statistical errors due to the correlation of the vector meson mass
and coupling and the polarization function. Both advantageous features of this mechanism have been put to work in \cite{Burger:2015lqa} as well with the above described positive outcome. Moreover, 
the direct comparison of the light-quark contribution from extrapolation in $\mps^2 \to \mpi^2$ and direct calculation at the physical pion mass, which is done without any modification of the observable,
provides further confirmation of the present results' consistency, but this
test has to be made more stringent by achieving higher statistical accuracy for the result from direct simulation at the physical pion mass.

The more accurate ab-initio determination of $\aleptonhlo$ directly at the physical pion mass is an on-going effort within the ETMC. The evaluation of the crucial light quark contribution for $N_f=2$ dynamical
fermions at physical pion mass along the lines of \cite{Abdel-Rehim:2015pwa} for a larger volume with $\mps\cdot L = 4$ is on-going. Moreover, with the continuing simulation activity of the ETMC aiming
for gauge field ensembles with fully dynamical up, down, strange and charm quark with physical values for the pion, Kaon and $D$-meson mass we have for the near future the prospect of calculating $\aleptonhlo$ in a
setup including QCD-dynamics from the first two families of quark flavors with quark masses at their physical values.

We are then basically left with the systematic effect of the finite volume in the form of the infrared cut-off $\sim 1/L$ of the lattice data. Lattice simulations in larger physical volumes
are an essential yet at present computationally cost-intensive way to constrain further the polarization function at smallest momenta. Beyond that alternative methods have been proposed and are being developed
(e.g. \cite{Bernecker:2011gh,Feng:2013xsa,Chakraborty:2015aaa}) to exploit further the analytic structure of the polarization function in finite volume to constrain its shape in the small momentum region.

\paragraph{Acknowledgments}
We thank all members of the organizing committee for organizing this highly informative and stimulating conference.
M.P. is grateful for the opportunity to attend the conference and present this work.
We are most grateful to Fred Jegerlehner for very enlightening discussions while carrying out these
computations. We thank the European Twisted Mass Collaboration for generating the gauge field ensembles used for the
calculations. This work has been supported in part by the DFG Corroborative Research Center
SFB/TR9. G.P. gratefully acknowledges the support of the German Academic National Foundation (Studienstiftung des deutschen Volkes e.V.) and of the DFG-funded Graduate School GK
1504.

\end{document}